\documentclass
[aps,twocolumn,preprintnumbers,amsmath,amssymb,superscriptaddress]{revtex4}%
\usepackage{amsfonts}
\usepackage{amsmath}
\usepackage{amssymb}
\usepackage{SIunits}
\usepackage{graphicx}
\usepackage{float}
\usepackage{textcomp}%
\providecommand{\U}[1]{\protect\rule{.1in}{.1in}}

\begin{document}

\title{Coherent terabit communications with microresonator Kerr frequency combs}
\author{Joerg~Pfeifle}
\affiliation{Institute of Photonics and Quantum Electronics (IPQ) and Institute of Microstructure Technology (IMT), Karlsruhe Institute of Technology (KIT), 76131, Karlsruhe, Germany}

\author{Victor~Brasch}
\affiliation{\'{E}cole Polytechnique F\'{e}d\'{e}rale de Lausanne (EPFL), 1015,
Lausanne, Switzerland}

\author{Matthias~Lauermann}
\affiliation{Institute of Photonics and Quantum Electronics (IPQ) and Institute of Microstructure Technology (IMT), Karlsruhe Institute of Technology (KIT), 76131, Karlsruhe, Germany}

\author{Yimin~Yu}
\affiliation{Institute of Photonics and Quantum Electronics (IPQ) and Institute of Microstructure Technology (IMT), Karlsruhe Institute of Technology (KIT), 76131, Karlsruhe, Germany}

\author{Daniel~Wegner}
\affiliation{Institute of Photonics and Quantum Electronics (IPQ) and Institute of Microstructure Technology (IMT), Karlsruhe Institute of Technology (KIT), 76131, Karlsruhe, Germany}

\author{Tobias~Herr}
\affiliation{\'{E}cole Polytechnique F\'{e}d\'{e}rale de Lausanne (EPFL), 1015,
Lausanne, Switzerland}

\author{Klaus~Hartinger}
\affiliation{Menlo Systems GmbH, 82152 Martinsried, Germany}

\author{Philipp~Schindler}
\affiliation{Institute of Photonics and Quantum Electronics (IPQ) and Institute of Microstructure Technology (IMT), Karlsruhe Institute of Technology (KIT), 76131, Karlsruhe, Germany}

\author{Jingshi~Li}
\affiliation{Institute of Photonics and Quantum Electronics (IPQ) and Institute of Microstructure Technology (IMT), Karlsruhe Institute of Technology (KIT), 76131, Karlsruhe, Germany}

\author{David~Hillerkuss}
\affiliation{Institute of Photonics and Quantum Electronics (IPQ) and Institute of Microstructure Technology (IMT), Karlsruhe Institute of Technology (KIT), 76131, Karlsruhe, Germany}
\affiliation{now with ETH Zurich, 8092 Zurich, Switzerland}

\author{Rene~Schmogrow}
\affiliation{Institute of Photonics and Quantum Electronics (IPQ) and Institute of Microstructure Technology (IMT), Karlsruhe Institute of Technology (KIT), 76131, Karlsruhe, Germany}

\author{Claudius~Weimann}
\affiliation{Institute of Photonics and Quantum Electronics (IPQ) and Institute of Microstructure Technology (IMT), Karlsruhe Institute of Technology (KIT), 76131, Karlsruhe, Germany}

\author{Ronald~Holzwarth}
\affiliation{Menlo Systems GmbH, 82152 Martinsried, Germany}

\author{Wolfgang~Freude}
\affiliation{Institute of Photonics and Quantum Electronics (IPQ) and Institute of Microstructure Technology (IMT), Karlsruhe Institute of Technology (KIT), 76131, Karlsruhe, Germany}

\author{Juerg~Leuthold}
\affiliation{Institute of Photonics and Quantum Electronics (IPQ) and Institute of Microstructure Technology (IMT), Karlsruhe Institute of Technology (KIT), 76131, Karlsruhe, Germany}
\affiliation{now with ETH Zurich, 8092 Zurich, Switzerland}

\author{T.~J.~Kippenberg}
\email{tobias.kippenberg@epfl.ch}
\affiliation{\'{E}cole Polytechnique F\'{e}d\'{e}rale de Lausanne (EPFL), 1015, Lausanne, Switzerland}

\author{Christian~Koos}
\email{christian.koos@kit.edu}
\affiliation{Institute of Photonics and Quantum Electronics (IPQ) and Institute of Microstructure Technology (IMT), Karlsruhe Institute of Technology (KIT), 76131, Karlsruhe, Germany}

\begin{abstract}
\bf{
Optical frequency combs enable coherent data transmission on hundreds of wavelength channels and have the potential to revolutionize terabit communications \citep{hillerkuss2011}. Generation of Kerr combs in nonlinear integrated microcavities \citep{haye2007} represents a particularly promising option enabling line spacings of tens of GHz, compliant with wavelength-division multiplexing (WDM) grids \citep{levy2012}. However, Kerr combs may exhibit strong phase noise and multiplet spectral lines \citep{herr2012,pfeifle2012a,wang2012}, and this has made high-speed data transmission impossible up to now. Recent work has shown that systematic adjustment of pump conditions enables low phase-noise Kerr combs with singlet spectral lines \citep{herr2012,li2012,Herr2013,okawachi2011,saha2013}. Here we demonstrate that Kerr combs are suited for coherent data transmission with advanced modulation formats that pose stringent requirements on the spectral purity of the optical source. In a first experiment, we encode a data stream of 392~Gbit/s on subsequent lines of a Kerr comb using quadrature phase shift keying (QPSK) and 16-state quadrature amplitude modulation (16QAM). A second experiment shows feedback-stabilization of a Kerr comb and transmission of a 1.44~Tbit/s data stream over a distance of up to 300 km. The results demonstrate that Kerr combs can meet the highly demanding requirements of multi-terabit/s coherent communications and thus offer a solution towards chip-scale terabit/s transceivers.}
\end{abstract}
\maketitle

Optical interconnects providing multi-terabit/s data rates are the most promising option to overcome transmission bottlenecks in warehouse-scale data centres and world-wide communication networks. By using highly parallel wavelength division multiplexing (WDM) with tens or hundreds of channels in combination with spectrally efficient advanced modulation formats, multi-terabit/s transmission capacity can be achieved while keeping symbol rates compliant with the electrical bandwidth of energy-efficient CMOS driver circuitry \citep{miller2009,qian2011}. The silicon platform allows for co-integration of photonic and electronic circuitry using fabless CMOS processing \citep{liu2010a,hochberg2010}. While integrated terabit/s WDM receivers have already been demonstrated \citep{feng2012}, scalability of the transmitter capacity is still limited by the lack of adequate optical sources, especially when using advanced modulation formats that encode information on both the amplitude and the phase of the optical wave and therefore require optical carriers with particularly low phase and amplitude noise. 

Optical carriers for WDM transmission are commonly generated by distributed feedback (DFB) laser arrays. Chip-scale transmitter systems with DFB lasers have been realized on indium phosphide (InP) substrates, showing potential for simultaneous operation of 40~channels \citep{nagarajan2010}. However, these approaches cannot be directly transferred to the silicon photonic platform: Combining conventional DFB laser arrays with silicon photonic transmitters would require a multitude of chip-chip interfaces, increasing significantly the packaging effort. Hybrid integration of III-V dies on silicon substrates \citep{park2005,fang2008_opex} avoids these interfaces, but scalability to large channel counts is still limited by the gain bandwidth of the semiconductor material and by thermal constraints. Moreover, spectral efficiency of DFB-based transmission systems suffers from the uncertainty of the individual emission frequencies, which is of the order of several GHz and requires appropriate guard bands to avoid spectral overlap of neighbouring WDM channels. For dense WDM with a channel spacing of, e.g., 25~GHz, the guard bands consume a significant fraction of the available transmission bandwidth.
 
\begin{figure*}
\includegraphics[width=1\textwidth]{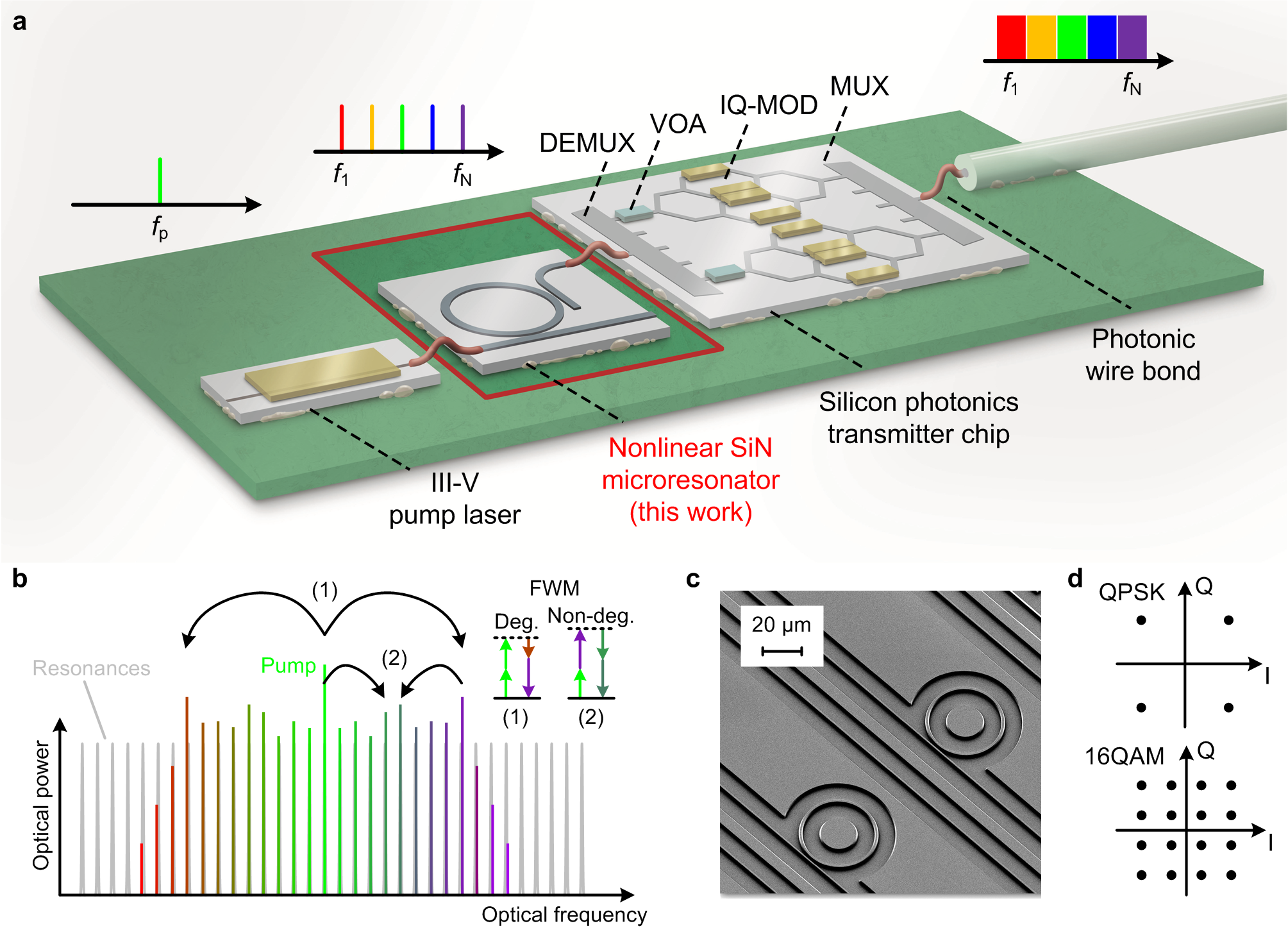}
\caption{\textbf{Principles of coherent terabit/s communications with Kerr frequency combs. a,} Artist's view of a possible future chip-scale terabit/s transmitter, leveraging a Kerr frequency comb source. For Kerr comb generation, a narrowband continuous-wave III-V laser pumps a high-Q silicon nitride microresonator. Resonant enhancement of the optical field in the presence of Kerr nonlinearity leads to formation of a frequency comb comprising a multitude of optical carriers at equidistant frequencies $f_1 ... f_{N}$; the associated spectra are depicted above the respective component. A WDM demultiplexer (DEMUX) separates the comb lines, and variable optical attenuators (VOA) are used to adjust the power of the individual carriers. Data are encoded on the in-phase and quadrature component by electro-optical modulators (IQ-MOD), using quadrature phase-shift keying (QPSK) or 16-state quadrature amplitude modulation (16QAM). The data streams are combined by a WDM multiplexer (MUX). Proper pulse shaping techniques enable a densely packed optical signal spectrum which is schematically depicted in the upper right. Photonic wire bonds \citep{lindenmann2012} connect the individual chips and link up the transmission fibre, thus enabling an efficient combination of different photonic integration platforms in a multi-chip assembly. The feasibility of the various silicon photonic passive components and electro-optic modulators has been shown \citep{liu2010a,dong2012a,korn2013}. The demonstration of coherent data transmission with Kerr combs is the subject of this work. \textbf{b,} Visual representation of the multi-stage four-wave mixing (FWM) process that leads to Kerr comb formation. Pump energy is transferred to the comb lines by two processes: Degenerate FWM, indicated as (1), converts two pump photons to a pair of photons that are up- and downshifted in frequency, whereas cascaded non-degenerate FWM processes, indicated as (2), fully populate the remaining resonances. \textbf{c,} Scanning-electron micrograph of an integrated high-Q SiN microresonator: High index-contrast SiN waveguides enable dense integration. \textbf{d,} Constellation diagrams of QPSK and 16QAM signals: Information is encoded both on the amplitude and the phase of the optical carrier, which can be represented by the in-phase (I, horizontal axis) and the quadrature component (Q, vertical axis) of the complex electrical field amplitude. Each point in the constellation diagram corresponds to a complex symbol that represents 2 (4) bits of information for the case of QPSK (16QAM).}
\label{Fig1}
\end{figure*}

These limitations can be overcome by exploiting optical frequency combs as sources for WDM transmission. Frequency combs consist of a multitude of equidistant spectral lines, each of which can be individually modulated \citep{hillerkuss2011,witzens2012,witzens2010a}. The inherently constant frequency spacing of the comb lines enables transmission of orthogonal frequency division multiplexing (OFDM) signals \citep{hillerkuss2011} or of Nyquist-WDM signals \citep{hillerkuss2012} with closely spaced subcarriers. Frequency combs with line spacings in the GHz range can be generated by external modulation of a narrowband continuous-wave signal \citep{wu2010}, by mode-locked lasers based on semiconductor quantum-dot or quantum-dash materials \citep{rosales2011}, or by exploiting parametric frequency conversion in Kerr-nonlinear high-Q microcavities \citep{haye2007}. In contrast to modulator-based approaches or mode-locked laser diodes, the bandwidth of Kerr combs is neither limited by the achievable modulation depth nor by the gain bandwidth of the active medium. Kerr combs can hence exhibit bandwidths of hundreds of nanometers \citep{kippenberg2011}, thereby covering multiple telecommunication bands (such as the C, L, and U band) with typical line spacings between 10~GHz and 100~GHz.

Kerr comb generation has been demonstrated using various different technology platforms \citep{kippenberg2011} such as silica \citep{haye2007}, calcium fluoride (CaF$_2$) \citep{savchenkov2008a}, Hydex glass \citep{razzari2010} or silicon nitride (Si$_3$N$_4$) \citep{levy2010}. Previous experiments have used such devices for data transmission with conventional 10~Gbit/s or 40~Gbit/s on-off-keying as a modulation format \citep{pfeifle2012a,wang2012,levy2012}. However, Kerr frequency combs tend to exhibit multiplet spectral lines within a single resonance leading to strong amplitude fluctuations and phase noise \citep{herr2012,pfeifle2012a}. Such fluctuations are prohibitive for spectrally efficient data transmission with advanced modulation formats. While low phase-noise Kerr combs have been demonstrated recently \citep{herr2012,li2012,Herr2013,okawachi2011,saha2013}, coherent transmission with Kerr combs has not yet been shown.

Here we report on the first experimental demonstration of coherent data transmission with amplitude- and phase-modulated carriers derived from a Kerr frequency comb \citep{pfeifle2013}. Coherent transmission allows to increase the information content and to boost the data rate, but also places stringent requirements on the phase and amplitude stability of the optical carrier. Our experiments build on systematic investigations of comb formation dynamics \citep{herr2012} to generate highly stable Kerr combs with low phase noise. We encode uncorrelated data on neighbouring comb lines using quadrature phase shift keying (QPSK) and 16-state quadrature amplitude modulation (16QAM) in combination with Nyquist pulses that have nearly rectangular power spectra and enable highest spectral efficiency. In a first experiment, we use polarization multiplexing and a symbol rate of 14~GBd on five QPSK and one 16QAM channel to obtain an aggregate data rate of 392~Gbit/s. In a second experiment \citep{pfeifle2014}, we boost the data rate to 1.44 Tbit/s, encoded on 20 neighbouring comb lines and transmitted over a distance of 300 km. The comb is stabilized by a feedback loop that controls the pump wavelength. This corresponds to a net spectral efficiency of 6~bit/s/Hz (3~bit/s/Hz) for the 16QAM (QPSK) channels. The results clearly demonstrate the large potential of Kerr frequency combs for future chip-scale terabit/s communication systems. We choose silicon nitride as a reliable integration platform which is compatible with CMOS processing \citep{levy2010,foster2011}.

\begin{figure*}
\includegraphics[width=1\textwidth]{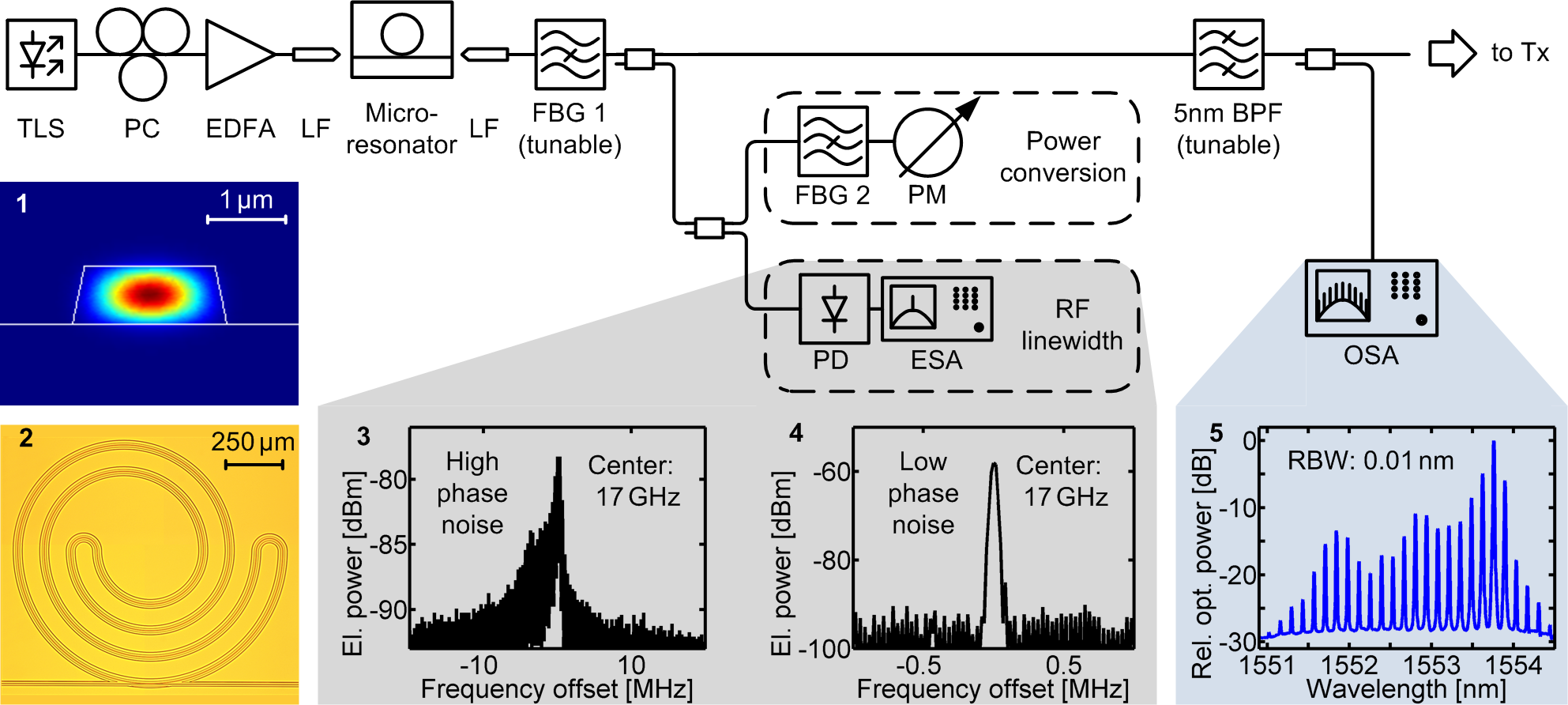}
\caption{\textbf{Comb-generation setup.} The optical pump comprises a tunable laser source (TLS), a polarization controller (PC), and an erbium-doped fibre amplifier (EDFA). Lensed fibres (LF) are used to couple light to and from the microresonator chip. The microresonator consists of a Si$_3$N$_4$ waveguide which is approximately 2~\micro m wide and 750~nm high with sidewalls inclined by 12$^\circ$ with respect to the vertical direction. The waveguide cross section and the mode profile (for a bend radius of 1.35~mm) are depicted in Inset 1. The ring resonator has a spiral-like layout to decrease the footprint and to reduce stitching errors during electron beam lithography, see Inset 2. After the microresonator, a fibre Bragg grating (FBG 1) serves as a tunable narrowband notch filter to suppress the residual pump light. For adjustment of the pump parameters, we monitor the power conversion from the pump to the adjacent lines. An electronic spectrum analyser (ESA) is used to measure the RF linewidth in the photocurrent spectrum of the photodetector (PD). When tuning into a low phase-noise comb state, the RF spectrum switches from a broad band (Inset 3, RBW 10~kHz) to a single narrow line (Inset 4, RBW 30~kHz). A 5~nm wide spectral section is extracted from the comb spectrum and used for data transmission (Inset 5). The pump wavelength and on-chip pump power are 1549.4~nm and 33~dBm, respectively.}
\label{Fig2}
\end{figure*}
 
The vision of a future chip-scale terabit/s transmitter is illustrated in Fig.~\ref{Fig1}a. A Kerr frequency comb is generated by exploiting multi-stage four-wave mixing (FWM) in a high-Q Kerr-nonlinear microresonator that is pumped with a strong continuous-wave (cw) laser \citep{haye2007,kippenberg2011}. The envisaged transmitter consists of a multi-chip assembly, where single-mode photonic wire bonds \citep{lindenmann2012} connect the individual chips. In contrast to monolithic integration this hybrid approach allows to combine the advantages of different photonic integration platforms: For the optical pump, III-V semiconductors can be used \citep{nagarajan2010}, while the high-Q ring resonator for Kerr comb generation could be fabricated using, e.~g., low-loss silicon-nitride waveguides \citep{levy2010}. The optical carriers are separated and individually modulated on a silicon photonic chip, for which large-scale silicon photonic integration lends itself to realize particularly compact and energy-efficient MUX and DEMUX filters \citep{liu2010a} and IQ modulators \citep{dong2012a,korn2013}.

Fig.~\ref{Fig1}b illustrates the basic principle of Kerr comb generation. Pump energy is transferred to the comb lines by two processes: Degenerate FWM, indicated as (1) in Fig. 1b, leads to formation of two sidemodes by converting two pump photons to a pair of photons that are up- and downshifted in frequency. The magnitude of the frequency shift is determined by pump power and cavity dispersion \citep{herr2012}. A multitude of cascaded non-degenerate FWM processes, indicated as (2) in Fig. 1b, fully populates the remaining resonances. Kerr comb generators can be extremely compact – Fig.~\ref{Fig1}c shows a scanning electron microscope picture of a planar integrated SiN microresonator. 

To maximize spectral efficiency, we use advanced modulation formats that encode data both on the amplitude and the phase of each comb line. This is illustrated in Fig.~\ref{Fig1}d for QPSK and 16QAM: The complex electric field of the data signal is visualized by its in-phase (I, horizontal axis) and quadrature component (Q, vertical axis) in the complex plane. For QPSK, the I and the Q-component can assume two distinct values, leading to four signal states (symbols) in the complex plane. This represents an information content of two bits per symbol. Likewise, a 16QAM symbol can assume 16 states in the complex plane, corresponding to 4 bits of information per symbol. A densely packed optical signal spectrum as schematically depicted in the upper right corner of Fig. 1a can be achieved by using sinc-shaped Nyquist pulses with rectangular power spectra \citep{schmogrow2012}.

\begin{figure*}
\includegraphics[width=1\textwidth]{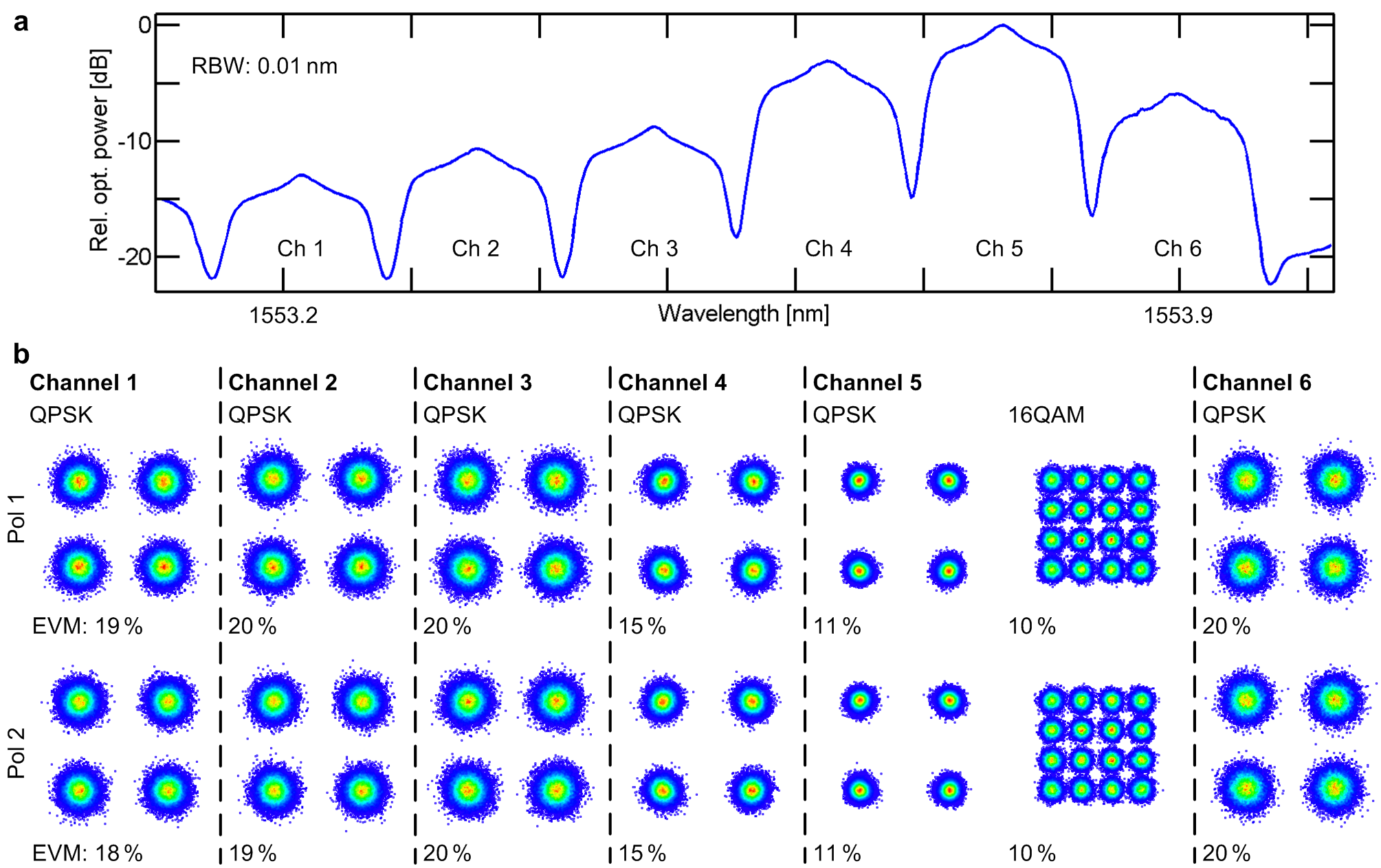}

\caption{\textbf{Coherent data transmission using a Kerr microresonator frequency comb. a,} Spectrum of modulated carriers for all six data channels, measured at the input of the optical modulation analyser (OMA). \textbf{b,} Constellation diagrams for each channel and for both polarizations along with the corresponding error vector magnitude (EVM) and the optical carrier-to-noise-density ratio (OCN$_0$R). The constellation diagrams show no sign of excessive phase noise, which would result in constellation points that are elongated along the azimuthal direction. For QPSK the BER of all channels is below $4.5\times10^{-3}$, which corresponds to an EVM of 38~\%; for channels 4 and 5 the BER is even smaller than $10^{-9}$ (EVM $<$ 17~\%). The good quality of channel 5 enables transmission of a 16QAM signal with a measured BER of $7.5\times10^{-4}$. }
\label{Fig3}
\end{figure*}

The viability of the concept illustrated in Fig.~\ref{Fig1} is demonstrated in a proof-of-principle experiment with discrete photonic components. Kerr combs are generated with the setup depicted in Fig.~\ref{Fig2}. Pump light from a narrow-linewidth tunable laser source (TLS) is adjusted in polarization and amplified by an erbium-doped fibre amplifier (EDFA). Lensed fibres (LF) couple the light to a Si$_3$N$_4$ ring resonator with a free spectral range (FSR) of 17~GHz, Fig.~\ref{Fig2}. Inset 1 shows the waveguide cross section of the ring and the calculated mode profile. The ring resonator waveguide is coiled up to reduce the footprint and thereby the stitching errors during electron beam lithography (Fig.~\ref{Fig2}, Inset 2), leading to a loaded Q-factor of  $8\times10^5$. In contrast to the concept illustrated in Fig.~\ref{Fig1}, our experiment, relies on a comb generator with a single waveguide which is used both to couple pump light to the resonator and to extract the frequency comb. As a consequence, strong cw pump light can pass through the resonator and needs to be suppressed by a tunable fibre Bragg grating (FBG 1) at the output of the device. It has recently been shown that stability and phase noise of the Kerr comb are closely linked to the pump conditions \citep{herr2012}. Careful tuning of pump power, frequency, and polarization is therefore of prime importance. To adjust these parameters in the experiment we used two optimization criteria: First, we measure the power conversion from the pump to all newly generated comb lines using a power meter (PM) after a second fibre Bragg grating (FBG~2) that completely blocks the pump. Second, we record the radio-frequency (RF) power spectrum of the photocurrent that is generated by direct detection of the comb. If the pump parameters are correctly adjusted, a single narrowband line is found in the photodetector current spectrum, indicating that the comb consists of equidistant narrowband spectral lines, see Fig.~\ref{Fig2}, Inset 4. Otherwise, if the lines are not frequency-locked, or if multiplet spectra exist, a broadband RF spectrum is observed, (Fig.~\ref{Fig2}, Inset 3). Further details of the adjustment procedure can be found in the Methods Section.

For data transmission, a 5~nm wide spectral section (Fig.~\ref{Fig2}, Inset 5) is extracted from the comb using an optical band-pass filter (BPF). The carriers are modulated with QPSK and 16QAM signals at a symbol rate of 14~GBd. To enable dense packing of optical channels and to maximize the spectral efficiency, we used sinc-shaped Nyquist pulses with rectangular power spectra \citep{schmogrow2012}. We transmit data streams on two orthogonal polarizations of a standard single-mode fibre (polarization-division multiplexing). The signal is detected with a commercial optical modulation analyser using a tunable laser as a local oscillator. The experimental setup for data transmission and the digital post-processing techniques are described in more detail in the Supplementary Information (SI).

The results of the data transmission experiment are summarized in Fig.~\ref{Fig3}. Fig.~\ref{Fig3}a depicts the optical power spectrum of the modulated carriers for all six data channels. We did not flatten the comb spectrum prior to modulation for comparing the influence of different carrier powers on the transmission performance. As a quantitative measure of the signal quality, we use the error vector magnitude (EVM), which describes the effective distance of a received complex symbol from its ideal position in the constellation diagram. The EVM is directly connected to the bit-error ratio (BER) if the signal is impaired by additive white Gaussian noise only \citep{schmogrow2012a}. The constellation diagrams for each polarization of the six wavelength channels are depicted in Fig.~\ref{Fig3}c along with the measured EVM. When using second-generation forward-error correction (FEC) with 7~\% overhead, the BER limits for error-free detection are given by $4.5\times10^{-3}$ \citep{chang2010}. For QPSK, this corresponds to an EVM threshold of 38~\%, which is well above the measured EVM for all channels, indicating that error-free transmission is possible. For 16QAM, requirements are more stringent, and an EVM of 11~\% is needed for error-free transmission. This is fulfilled for channel 5, which shows a measured BER of $7.5\times10^4$. We transmit with a symbol rate of 14~GBd and choose QPSK for channels 1...4 and 6, and 16QAM for channel 5. Taking into account polarization multiplexing, we obtain an aggregate data rate of 392~Gbit/s. Considering the overhead of 7~\% for second generation FEC, the net spectral efficiency amounts to 3~bit/s/Hz for the QPSK and to 6~bit/s/Hz for the 16QAM channels.

\begin{figure*}
\includegraphics[width=1\textwidth]{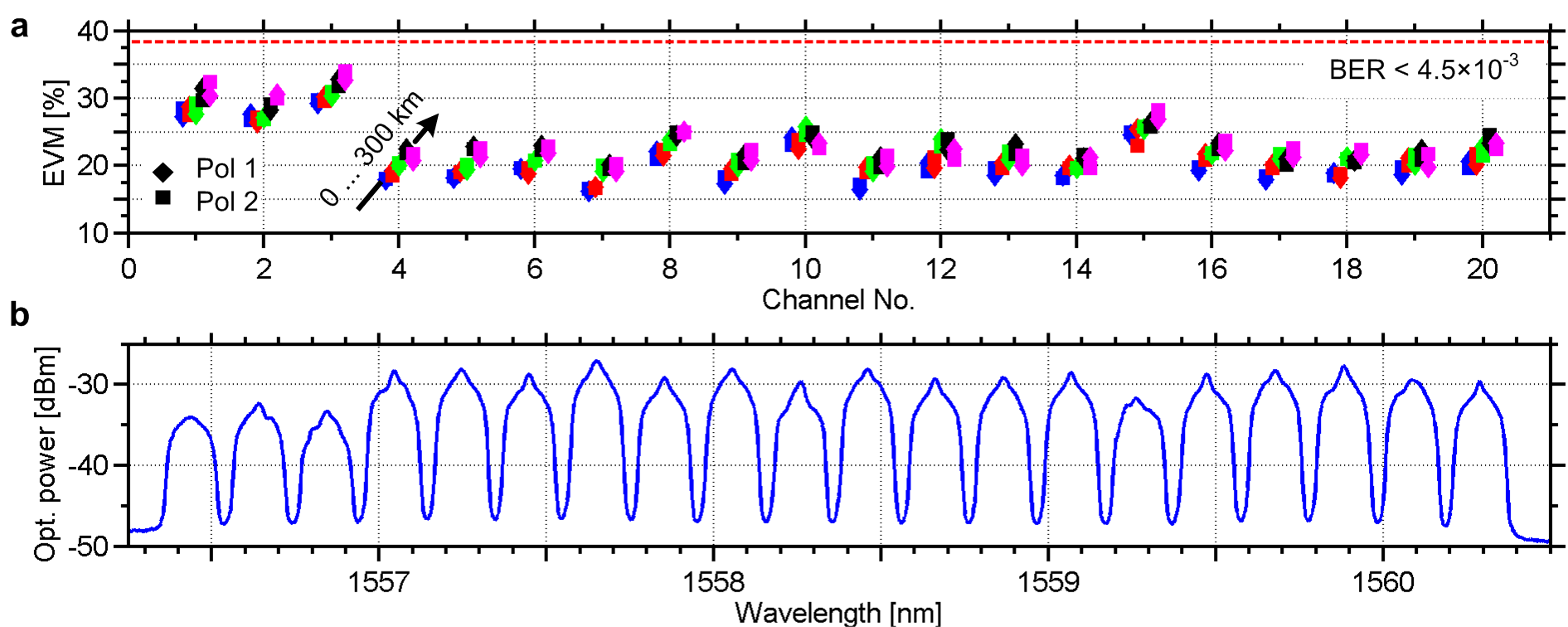}

\caption{\textbf{Coherent terabit/s data transmission using a feedback-stabilized Kerr frequency comb. a,} EVM for all data channels and fibre spans. The carriers are modulated at a symbol rate of 18~GBd using QPSK and Nyquist pulse shaping. With polarization multiplexing at each of the 20~WDM channels an aggregate data rate of 1.44~Tbit/s is achieved. The polarizations are distinguished by diamonds and squares, while the different fibre spans (0~km, 75~km, 150~km, 225~km, 300~km) are color-coded and slightly offset in the horizontal direction as indicated by the arrow. The red dashed line indicates an EVM of 38~\%, which corresponds to the BER threshold for second-generation FEC. \textbf{b,} Optical spectrum of the 1.44~Tbit/s data stream. The spectrum was flattened prior to modulation. }
\label{Fig4}
\end{figure*}

When relating the EVM to the BER, we assume that the signal is impaired by additive white Gaussian noise. The validity of this assumption is supported by the fact that the deviations of the measured from the ideal constellation points occur equally in all directions and do not show any sign of anisotropy. Excessive phase noise, in particular, can be excluded as a relevant impairment of data transmission as this would lead to constellation points which are elongated along the azimuthal direction. The results indicate that Kerr combs are indeed perfectly suited for data transmission with phase-sensitive modulation formats.

The remaining signal impairments can be attributed to strong amplified spontaneous emission (ASE) originating from the high-power pump EDFA. In the current configuration, the ASE light passes straight through the resonator chip and superimposes the comparatively weak comb lines as additive white Gaussian noise. The power of the comb lines is limited by the efficiency of the frequency conversion and the coupling of the pump to the resonator. As a figure of merit we define the optical carrier-to-noise-density ratio (OCN$_0$R), which relates the power of the unmodulated carrier to the underlying noise power in a spectral bandwidth of 1~Hz, see Methods Section for more details. In our experiment, the OCN$_0$R is correlated to the EVM: For channel 5, we obtain a good OCN$_0$R of more than 119~dB~Hz, which enables a very good QPSK signal and a good 16QAM signal. Channel 4 with an OCN$_0$R of more than 115~dB~Hz still gives a very good QPSK signal but lacks performance for 16QAM. Channels 1...3 as well as channel 6 exhibit even lower OCN$_0$R resulting in larger EVM figures.

In a second experiment, we use a resonator with a FSR of 25~GHz and a Q-factor of $2\times10^{6}$, which is higher than in the first experiment. This allows reducing the pump power to a level where filtering of the ASE noise from the EDFA is possible. To enable stable long-term operation of the Kerr comb for extended data transmission experiments, we implement a feedback loop, which locks the wavelength of the pump laser to a specified position within the resonance of the cavity, see Methods Section and SI for more details. In contrast to the previous experiment, we now flatten the comb lines prior to modulation using a programmable filter. For data transmission we use QPSK at a symbol rate of 18~GBd, and we insert up to four fibre spans of 75~km length between the transmitter and the receiver. The results are summarized in Fig.~\ref{Fig4}. We obtain 20 channels with an EVM below the threshold of 38~\%, that corresponds to a BER of $4.5\times10^{-3}$. The total data stream amounts to 1.44~Tbit/s with a net spectral efficiency of 2.7~bit/s/Hz. This is the highest data rate that has ever been transmitted using a Kerr comb as WDM source.

In summary we demonstrate that Kerr frequency combs are well suited for high-capacity data transmission with phase-sensitive modulation formats. We show error-free transmission with data rates of up to 1.44~Tbit/s, spectral efficiencies of up to 6~bit/s/Hz, and transmission distances of up to 300~km. The received signals exhibit no sign of excessive phase noise. Assuming that the demonstrated spectral efficiency of 6~bit/s/Hz can be maintained over the entire bandwidth of the Kerr comb, we envision chip-scale transmitters providing aggregate data rates beyond 100~Tbit/s, only limited by nonlinear effects in the single-mode silica fibres used for transmission \citep{essiambre2010}. In the long run, these limitations might even be overcome by mode division multiplexing based on, e.g., exploiting the orbital angular momentum (OAM) of the fiber modes to create orthogonal data streams \citep{bozinovic2013}. The combination of chip-scale Kerr frequency comb sources with large-scale silicon photonic integration could hence become a key concept for power-efficient optical interconnects providing transmission rates that have hitherto been considered impossible.

\section*{Acknowledgemets}
This work was supported by the European Research Council (ERC Starting Grant 'EnTeraPIC', number 280145), by the Alfried Krupp von Bohlen und Halbach Foundation, by Helmholtz International Research School for Teratronics (HIRST), by the Initiative and Networking Fund of the Helmholtz Association, by the Center for Functional Nanostructures (CFN) of the Deutsche Forschungsgemeinschaft (DFG) (project A 4.8), by the DFG Major Research Instrumentation Programme, by the Karlsruhe Nano-Micro Facility (KNMF), by the Karlsruhe School of Optics \& Photonics (KSOP), by the Swiss National Science Foundation (NCCR Nano-Tera, NTF MCOMB), by a Marie Curie IAPP Action and by the DARPA program QuASAR. Samples have been fabricated at the EPFL Center for Micro- and Nanotechnology (CMI).

\section*{Methods}

\subsection*{Experimental setup and ring resonator design.} 
The pump for Kerr comb generation is generated by an external-cavity laser (New Focus Velocity Model TLB-6728), a polarization controller (PC) and an erbium-doped fibre amplifier (EDFA) providing an output power of up to 37~dBm, see Fig.~\ref{Fig2}. The coupling loss between the lensed fibre and the Si$_3$N$_4$ chip amounts to approximately 3~dB per facet. In the data transmission experiments we pump a resonance near 1549.4~nm. The tunable fibre Bragg grating (FBG~1) at the output of the device suppresses the remaining pump signal by approximately 20~dB. The ring resonator consists of nearly stoichiometric Si$_3$N$_4$ grown in multiple layers with intermediate annealing steps \citep{levy2010}. The strip waveguides are patterned with electron beam lithography and transferred to the substrate by reactive ion etching with SF$_6$/CH$_4$ chemistry. The waveguides are then embedded into a SiO$_2$ cladding using low-pressure chemical vapour deposition (LPCVD). The waveguides are 2~\micro m wide and 750~nm high, and the sidewalls are inclined by 12$^\circ$ with respect to the vertical direction (Fig.~\ref{Fig2}, Inset 1). 

The high-temperature growth technique used to fabricate the 750~nm thick layers of stoichiometric Si$_3$N$_4$ requires annealing at 1200$^\circ$C. This makes it difficult to fabricate SiN resonators in the framework of standard CMOS processes. One option to overcome these limitations is to use dedicated fabrication processes and multi-chip integration as illustrated in Fig.~\ref{Fig1}. Alternatively, it is possible is to deposit Si$_3$N$_4$ at 400$^\circ$C and to use UV thermal processing (UVTP) at lower temperatures to reduce the defect density. Both techniques are subject to ongoing research.

In the first data transmission experiment, the resonator exhibits a free spectral range (FSR) of 17~GHz, and a loaded Q-factor of $8\times10^5$. Kerr combs are generated using an on-chip pump power of approximately 33~dBm. For the second experiment, we fabricated a resonator with a FSR of 25~GHz, which exhibits an increased Q-factor of $2\times10^{6}$. The comb is generated with approximately 29~dBm of on-chip pump power, and the measured line spacing deviates by less than 6~MHz from the designed 25~GHz. We expect that microresonators with further improved Q-factors will enable generation of broadband frequency combs with hundreds of lines that cover optical bandwidths of hundreds of nanometers - a multiple of what can be achieved with state-of-the-art distributed-feedback WDM laser arrays in InGaAsP technology. Note that in the envisaged system illustrated in Fig.~\ref{Fig1}, the use of FBG for pump light suppression will become superfluous: If the frequency comb is extracted by a second waveguide coupled to the ring resonator, direct through-coupling of strong cw pump light is not possible, and FBG notch filters are not needed.

\subsection*{Generation of low phase-noise Kerr combs.}
To obtain low phase-noise Kerr combs, the pump parameters are adjusted in two steps using the setup depicted in Fig.~\ref{Fig2}: First, the pump wavelength is periodically scanned across the resonance at a frequency of approximately 100~Hz while staying within the stop band of FBG~2 and continuously measuring the power conversion to the spectral region outside the stop band. During these scans the polarization of the pump signal is slowly varied to maximize the conversion. Maintaining this polarization, the detuning of the pump signal with respect to the resonance wavelength is then carefully adjusted until the initially broad RF spectrum (Fig.~\ref{Fig2}, Inset 3) exhibits a single narrow peak (Fig.~\ref{Fig2}, Inset 4). Note that the frequency axes of Insets 3 and 4 have different scales. The power spectrum of Inset 3 was recorded with a resolution bandwidth (RBW) of 10~kHz whereas Inset 4 was recorded with a RBW of 30~kHz. For an on-chip pump power of approximately 33~dBm the essential part of the optical comb spectrum is depicted in Inset 5 of Fig.~\ref{Fig2}. From this partial spectrum we selected the lines for our data transmission experiments. Spectra of the comb used for the second experiment can be found in the SI. 

The on-chip pump power needed for Kerr comb generation in the first and in the second experiment, still amounts to 33~dBm and 29~dBm, respectively. However, there is considerable room for future improvements - Kerr combs have been demonstrated with a threshold power as low as 50~\micro W in silica toroids \citep{haye2007}. In general, the threshold for Kerr comb generation scales inversely quadratic with the Q-factor of the resonator \citep{kippenberg2004a}. A 10-fold Q-factor improvement would lead to a reduction of the pump power by a factor of 100 and the 2~Watt pump laser could be replaced by 20~mW pump laser diode.

\subsection*{Feedback stabilization of a Kerr comb. }
To maintain low-phase noise comb states during an extended data transmission experiment, it is important to keep the pump conditions as constant as possible. In the first experiment, we found that the on-chip pump power is a crucial parameter and must be kept constant to approximately $\pm$5~\% to maintain thermal locking of the cavity resonance to the pump wavelength \citep{carmon2004}. The second experiment relies on stable comb generation for extended studies of data transmission performance. This is achieved by an independent control loop based on commercially available hardware (Melles Griot NanoTrak) that keeps the lensed fibres in the optimum coupling position. In addition, we implemented a second control loop using a commercially available PID controller (TEM Messtechnik LaseLock), which takes the power in the comb lines as feedback signal and controls the pump laser detuning such that thermal drifts of the resonator are followed by slight adaptation of the pump wavelength, see SI for a more detailed discussion. The combination of these two control loops enables stable Kerr comb operation over several hours without human interaction.

\begin{figure*}
\includegraphics[width=1\textwidth]{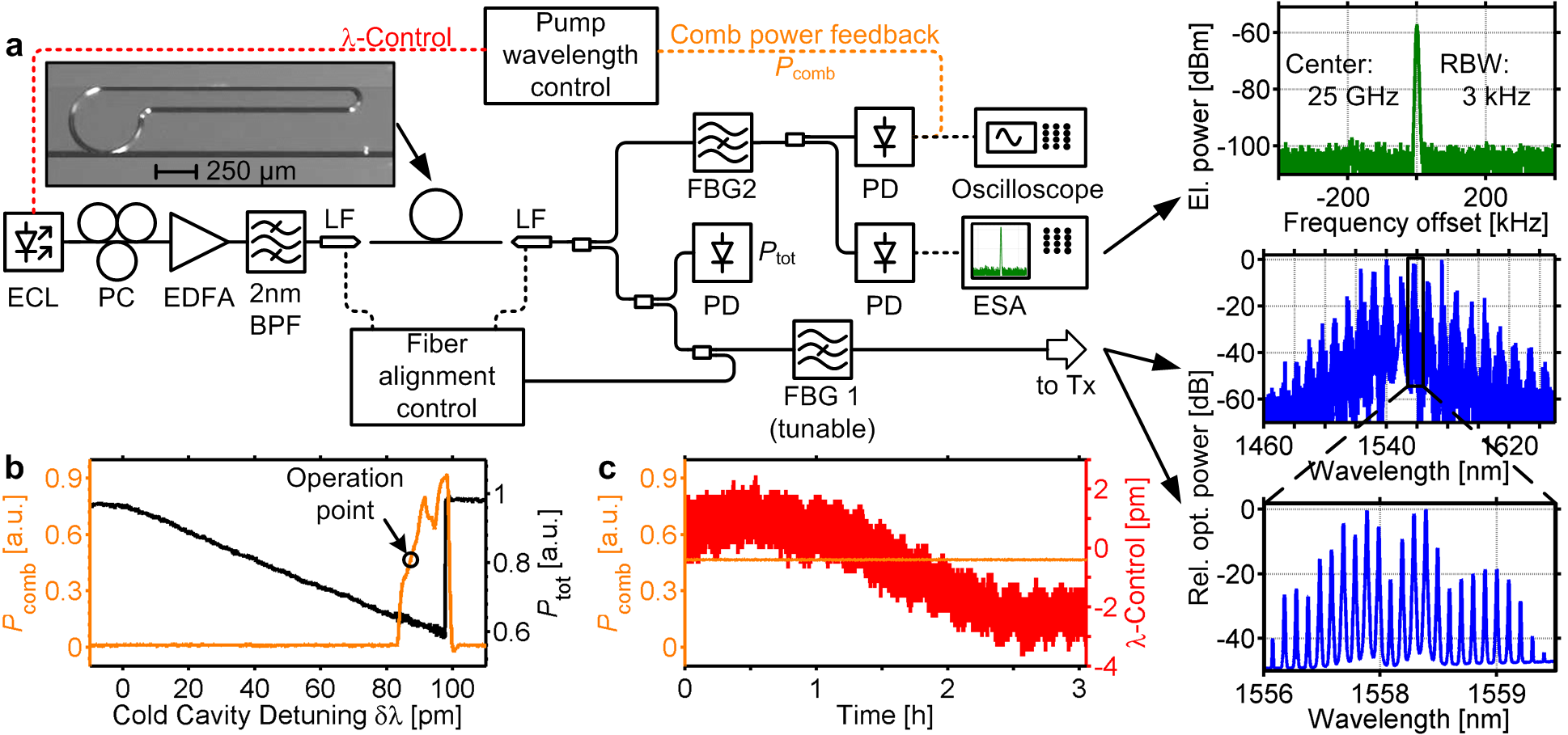}

\caption{\textbf{Generation and stabilization of a Kerr frequency comb in the second transmission experiment. a,} The experimental setup is similar to the one used in the first experiment, see Fig.~\ref{Fig2}. Due to an improved Q-factor, the pump power could be reduced which allows filtering of the ASE noise originating from the pump EDFA using a 2~nm band pass filter (BPF). The resonator (inset 1) exhibits a FSR of 25~GHz and a loaded Q-factor of $2\times10^{6}$. Using an on-chip power of 29~dBm, we pump a resonance near 1549.4~nm. The pump parameters are manually adjusted using the scheme described in the Methods Section. The insets on the right-hand side show the RF (top) and optical (mid and bottom) spectra of the resulting comb. For a stable comb output, we control the wavelength of the pump laser. As a feedback signal, we use the comb power $P_{comb}$ measured behind FBG~1. \textbf{b,} $P_{comb}$ (left axis, orange) and the total power $P_{tot}$ measured at the output of the chip (right axis, black) as a function of the detuning $\delta\lambda$ from the cold cavity resonance. The circle indicates the operation point of the control loop, which corresponds to the manually adjusted low phase-noise comb state. The control loop is activated after the pump parameters have been optimized. \textbf{c,} Feedback and wavelength control signal for a time frame of three hours. The pump wavelength follows the thermal drifts of the resonance, which leads to perfectly stable comb output power.}
\label{Fig5}
\end{figure*}

\subsection*{Signal impairment analysis. } 
In the first experiment, the optical carrier-to-noise-density ratio (OCN$_0$R) is limited by the ASE noise originating from the pump amplifier. To determine the OCN$_0$R, we need to estimate the noise power density right at the frequency of the respective comb line. This is done by averaging the spectral power densities obtained in the middle between two comb lines. The spectral power densities were measured with an optical spectrum analyser (Ando AQ 6317B) in a 0.01~nm bandwidth and renormalized to a bandwidth of 1~Hz.

\begin{figure*}
\includegraphics[width=1\textwidth]{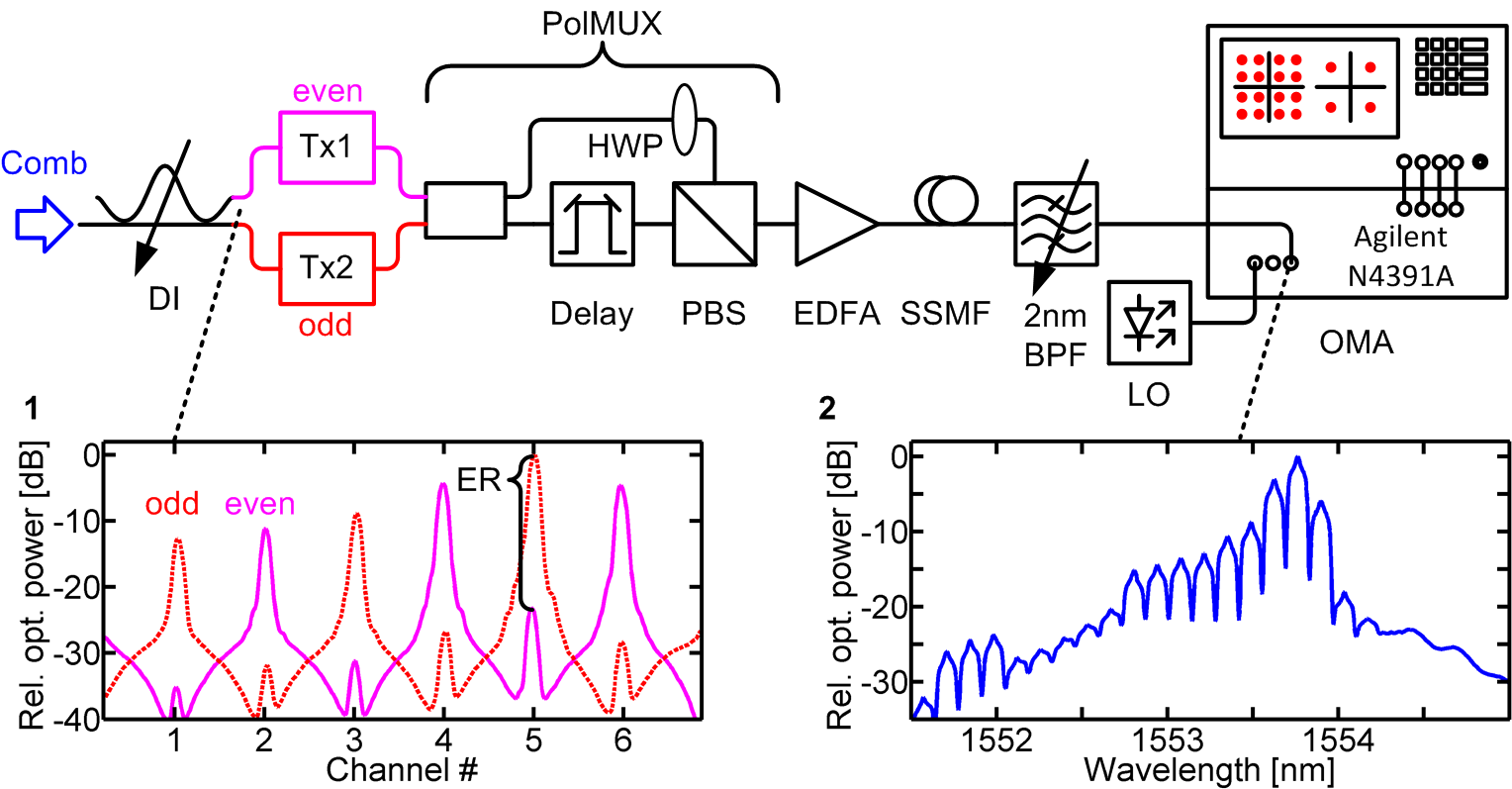}

\caption{\textbf{Data transmission setup.} A tunable delay interferometer (DI) is used to separate the incoming frequency comb into carriers that belong to odd and even channels. The two sets of channels are then separately modulated at a symbol rate of 14~GBd using QPSK or 16QAM signals and Nyquist sinc-pulses. The two data streams are merged and fed into a polarization multiplexer (PolMUX), comprising a half-wave plate (HWP), a delay line and a polarization beam splitter (PBS). The PolMUX splits the optical signal in two parts, delays one of them to form two uncorrelated data streams, and recombines them on two orthogonal polarization states of the standard single-mode transmission fibre (SSMF). The receiver comprises an erbium-doped fibre amplifier (EDFA), an optical band-pass filter (BPF), and an optical modulation analyzer (OMA) with a tunable laser serving as a local oscillator (LO). Digital signal processing in the baseband is used to separate the WDM channels and to demultiplex the polarizations. Insets: \textbf{1,} Spectra of the unmodulated carriers after separation by the DI. Even channels are represented by the magenta solid line, odd channels by the red dotted line. The extinction ratio is better than 20~dB for all channels; \textbf{2,} Optical spectrum at the input of the OMA, after the 2~nm band-pass filter. The pronounced noise background predominantly originates from the ASE of the pump EDFA that is used for frequency comb generation. }
\label{Fig6}
\end{figure*}

To improve the signal quality, it is possible to filter out ASE by using a band-pass filter before the DUT as demonstrated in the second experiment. When performing the first experiment, however, we did not have any filters that could safely handle the 37~dBm of pump power right after the EDFA. We showed that by further optimizing the fabrication process, we can increase the microresonator's Q-factor and hence significantly decrease the required threshold power \citep{kippenberg2004a}, such that filtering of the pump signal was possible. In addition the comb may be extracted by a second waveguide which is coupled to the microresonator, see Fig.~\ref{Fig1}. This avoids direct transmission of broadband ASE noise transmitted through the resonator device. Using resonators with higher Q-factor, it might even be possible to use a high-power pump laser diode without requiring amplification at all.

\section*{Supplementary Information}

\subsection*{Feedback-stabilization of a Kerr frequency comb}

To maintain low phase-noise comb states during the second experiment, we have implemented a control loop that takes the power $P_{comb}$ measured at the output of FBG~2 as a feedback signal and adjusts the wavelength of the pump laser to ensure a constant detuning. Fig.~\ref{Fig5}a shows the comb generation setup along with the control loop. The principle of the feedback stabilization is illustrated in Fig.~\ref{Fig5}b, which shows $P_{comb}$ (orange trace, left axis) and the total chip output power $P_{tot}$ (black trace, right axis) as a function of the detuning $\delta\lambda$ between the pump wavelength and the resonance wavelength of the cold cavity. As the wavelength of the pump is scanned, it drags along the resonance through thermal locking. This leads to a continuous decrease of $P_{tot}$ until locking is lost for large detuning. At the same time, $P_{comb}$ exhibits a pronounced increase once the power coupled to the cavity reaches the threshold for Kerr comb generation. Our control loop is based on a commercially available PID controller (TEM Messtechnik LaseLock) and operates at the rising edge of $P_{comb}$. The circle in Fig.~\ref{Fig5}b indicates the operation point that corresponds to the low phase-noise comb state for which the pump parameters have been manually optimized before starting the control. Moreover, for stable comb generation, it is crucial that also the fibre-chip coupling efficiency is actively stabilized. To this end, we use an independent control loop based on commercially available hardware (Melles Griot NanoTrak) to keep the lensed fibres in the optimum coupling position. We confirm the functionality of the feedback stabilization by recording the feedback signal and the comb power for more than 3h, Fig.~\ref{Fig5}c. The pump wavelength follows the thermal drifts of the resonance, and residual comb power fluctuations are found to be well below 1~\%. Comb spectra measured at the beginning and at the end of the extended data transmission experiment are indistinguishable within the accuracy of our spectrometer (ANDO AQ 6317B).

\subsection*{Experimental setup for WDM data transmission}

To emulate a WDM communication system, we transmit data on all channels simultaneously, using one pseudo-random bit sequence (PRBS) on the even channels and another independent PRBS on the odd channels. Neighbouring channels hence carry uncorrelated data streams, and inter-channel crosstalk translates directly into a degradation of signal quality. For the first data transmission experiment, the corresponding setup is depicted in Fig.~\ref{Fig6}. Carriers belonging to odd and even channels are first separated by a tunable delay interferometer (DI) \citep{li2011}. The superposition of both spectra is displayed in Inset 1 of Fig.~\ref{Fig6}. Two home-built multi-format transmitters \citep{schmogrow2010} are then used to encode QPSK or 16QAM signals with a symbol rate of 14~GBd on the sets of even and odd carriers using independent PRBS with a length of $2^{15} - 1$. Polarization-division multiplexing (PDM) is emulated by splitting the optical signal, and recombining the original and a sufficiently delayed copy to form two decorrelated data streams on orthogonal polarizations. The data stream is transmitted to the receiver, where the signal is amplified, filtered and coherently detected using an optical modulation analyser (OMA, Agilent N 4391 A) with a standard tunable laser (Agilent 81680A) acting as a local oscillator (LO). An optical bandpass filter with a 2~nm passband is used to avoid saturation of the receiver photodiodes. Individual WDM channels are selected by appropriate choice of the LO wavelength and by digital brick-wall filtering in the baseband. Inset 2 of Fig.~\ref{Fig6} shows the optical spectrum at the input of the OMA. The effect of the 2~nm band-pass filter is clearly visible; the noise in the pass-band predominantly originates from the ASE of the pump EDFA that is used for frequency comb generation. 

For the second data transmission experiment, the setup differs only slightly from the one depicted in Fig.~\ref{Fig6}: Instead of the delay interferometer, we use a programmable filter (Finisar WaveShaper) to separate the comb lines, which additionally allows flattening the spectral envelope of the comb. Moreover, we only use QPSK as a modulation format, and increase the symbol rate to 18~GBd. In addition to back-to-back testing, we transmit the signal over up to four spans of 75~km standard single mode fibre (SSMF) with an EDFA set to an output power of 10~dBm before each span.

In both transmission experiments, digital post-processing comprises demultiplexing of the polarization channels, compensation of the frequency offset between the LO and the optical carrier, clock recovery, equalization, and dispersion compensation where applicable. For performance evaluation we use the error-vector magnitude (EVM) metric, which describes the effective distance of a received complex symbol from its ideal position in the constellation diagram. Provided that the signal is impaired by additive white Gaussian noise only, the EVM is directly related \citep{schmogrow2012a} to the bit-error ratio (BER). For 16QAM, we estimate a BER of $3\times10^{-4}$ from the measured EVM of 10~\%, which is in fair agreement with the measured BER of $7.5\times10^{-4}$. This confirms the validity of the EVM-based estimation of the BER in our experiment. For the other QPSK channels, the BER of $3\times10^{-7}$, $7\times10^{-8}$, $1\times10^{-8}$, $1\times10^{-11}$, and $5\times10^{-20}$ are estimated accordingly from the measured EVM values of 20~\%, 19~\%, 18~\%, 15~\%, and 11~\%. Channels 4 and 5 hence feature BER, well below the threshold of $10^{-9}$, which is commonly considered error-free in telecommunications. This clearly demonstrates that error-free data transmission with Kerr combs is possible and that the achievable BER is rather limited by the components of the transmission system such as modulators, amplifiers and filters than by the coherence of the light source.

Apart from the ASE of the pump EDFA as discussed in the Methods Section, we investigated coherent crosstalk as a potential source of signal degradation: In the first experiment, the tunable DI used to separate the carriers has a finite extinction ratio (ER), see Inset 1 in Fig.~\ref{Fig6}. As a consequence, the data signals of the even channels are superimposed by residual carriers modulated with odd-channel data and vice versa, which leads to further signal degradation. However, in our experiment this effect was not relevant: With an ER better than ($22.2\pm1.5$)~dB for all channels, no correlation of channel ER and EVM can be observed. In the second experiment using the programmable filter we achieve an ER of ($24.6\pm4.0$)~dB. Coherent crosstalk can therefore safely be excluded as a source of signal degradation.

\newpage

\renewcommand{\emph}{\textit}
\bibliographystyle{naturemag}
\bibliography{Bibliography}

\end{document}